\documentclass[11pt,a4paper]{article}
\usepackage{jcappub}
\usepackage{bm}
\usepackage{amsmath}
\def\dd{\mathrm{d}}
\def\mcH{\mathcal{H}}
\def\mcP{\mathcal{P}}

\def\em{{\rm em}}

\def\rad{{\rm rad}}

\def\Mpl{M_{\rm Pl}}
\def\Mpc{{\rm Mpc}}
\def\GeV{{\rm GeV}}
\def\CMB{{\rm CMB}}

\def\G{{\rm G}}
\def\reh{{\rm reh}}

\newcommand{\HT}[1]{{\bf {\color{cyan} HT:  #1}}}

\def\osc{{\rm osc}}

\def\dec{{\rm dec}}

\def\bmB{\bm{B}}

\def\mcA{\mathcal{A}}

\title{Consistent generation of magnetic fields in axion inflation models}
\author[a,b]{Tomohiro Fujita,}
\author[a]{Ryo Namba,}
\author[a,b]{Yuichiro Tada,}
\author[c]{Naoyuki Takeda}
\author[d]{and Hiroyuki Tashiro}
\affiliation[a]{Kavli Institute for the Physics and Mathematics of the
Universe (Kavli IPMU), WPI, TODIAS,  the University of Tokyo, 5-1-5
Kashiwanoha, Kashiwa, 277-8583, Japan}
\affiliation[b]{Department of Physics, Graduate School of Science,
The University of Tokyo, Bunkyo-ku 113-0033, Japan}
\affiliation[c]{Institute for Cosmic Ray Research, University of Tokyo, Kashiwa, Chiba 277-8582, Japan}
\affiliation[d]{Department of Physics and Astrophysics, Nagoya University, Nagoya 464-8602, Japan}
\emailAdd{tomohiro.fujita@ipmu.jp}
\emailAdd{ryo.namba@ipmu.jp}
\emailAdd{yuichiro.tada@ipmu.jp}
\emailAdd{takedan@icrr.u-tokyo.ac.jp}
\emailAdd{hiroyuki.tashiro@nagoya-u.jp}

\abstract{
There has been a growing evidence for the existence of magnetic fields in the extra-galactic regions, while the attempt to associate their origin with the inflationary epoch alone has been found extremely challenging. We therefore take into account the consistent post-inflationary evolution of the magnetic fields that are originated from vacuum fluctuations during inflation. In the model of our interest, the electromagnetic (EM) field is coupled to a pseudo-scalar inflaton $\phi$ through the characteristic term $\phi F\tilde F$, breaking the conformal invariance. 
This interaction dynamically breaks the parity and
enables a continuous production of only one of the polarization states of the EM field through tachyonic instability. The produced magnetic fields are thus helical. We find that the dominant contribution to the observed magnetic fields in this model comes from the modes that leave the horizon near the end of inflation, further enhanced by the tachyonic instability right after the end of inflation. The EM field is subsequently amplified by parametric resonance during the period of inflaton oscillation. Once the thermal plasma is formed (reheating), the produced helical magnetic fields undergo a turbulent process called inverse cascade, which shifts their peak correlation scales from smaller to larger scales. We consistently take all these effects into account within the regime where the perturbation of $\phi$ is negligible and obtain $B_{\rm eff} \sim 10^{-19} \G$, indicating the necessity of additional mechanisms to accommodate the observations. 
}

\keywords{inflation, primordial magnetic fields}
\arxivnumber{1503.05802}

\begin{document}

\begin{flushright}
ICRR-Report-699-2014-25
\\
IPMU 15-0029
\end{flushright}

\maketitle

%
%
%
\section{Introduction}

Magnetic fields are observed on many different scales in our universe. In particular, it has been known that galaxies and galaxy clusters
have their own magnetic fields with the typical strength, $10^{-6}\G$~\cite{Wielebinski:2005, Bernet:2008qp, Beck:2009, Bonafede:2010xg, Feretti:2012vk}. 
Although their existence is observationally confirmed, the origin of these magnetic fields is still one of the open questions in cosmology.
Recently, the gamma-ray observations from blazars suggest the existence of
cosmological magnetic fields even in voids~\cite{Neronov:1900zz, Tavecchio:2010mk, Dolag:2010ni, Essey:2010nd, Taylor:2011bn, Takahashi:2013uoa, Finke:2013bua,Chen:2014rsa}.
They set the lower bound on the effective strength of the void magnetic fields, given as~\cite{Taylor:2011bn,Neronov:2009gh}
\begin{equation}
B_{\rm eff}  \gtrsim 10^{-15}\G,
\qquad
B_{\rm eff} \equiv B\times \left\{
\begin{array}{lc}
\sqrt{\lambda/1\Mpc} \quad & (\lambda<1\Mpc)\\
1 &(\lambda>1\Mpc)
\end{array}\right.,
\label{obs lower bound}
\end{equation}
where $B$ is the present field amplitude with the correlation length $\lambda$.%
\footnote{If one assumes that the suppression of the cascade emission is caused by the time delay of the cascade photons, 
one obtains a more conservative bound as $B_{\rm eff}\gtrsim
10^{-17}$~\cite{Taylor:2011bn}.}
The recent simulation indicates that if magnetic fields whose comoving strength is larger than $10^{-19}\G$ exist prior to galaxy formation, 
they can be amplified to $10^{-6}\G$ as the disk of galaxy develops~\cite{Pakmor:2013rqa}.
Therefore the reported magnetic fields in voids can seed those in
galaxies and galaxy clusters.
Hence ``primordial magnetogenesis'' can be a key ingredient of cosmology to explain the existence of magnetic fields in both galactic and extragalactic scales.
Magnetogenesis during inflation has attracted attention, and
many models have been proposed so
far~\cite{Turner:1987bw,Ratra:1991bn,Garretson:1992vt,Lemoine:1995vj,Gasperini:1995dh,Finelli:2000sh,Davis:2000zp,Bamba:2003av,Anber:2006xt,Martin:2007ue,Durrer:2010mq,Caprini:2014mja,Calzetta:2014eaa,Cheng:2014kga,Tasinato:2014fia}
(see also ref.~\cite{Kobayashi:2014sga}).  
In such scenarios, large-scale magnetic fields are expected to be generated
during inflation via its causal production of fluctuations.
Recently, however, the obstacles of inflationary magnetogenesis have been recognized,
and it is understood to be quite difficult to generate magnetic fields only
during inflation with the field strength satisfying the observational lower bound given by eq.~\eqref{obs lower bound}~\cite{Demozzi:2009fu,Barnaby:2012tk,Fujita:2012rb,Fujita:2013qxa,Fujita:2014sna,Ferreira:2014hma}.
 Since magnetic fields are substantially diluted after inflation due to
the cosmic expansion, 
their energy density needs to be extremely large during
inflation, altering the inflation dynamics and/or inducing an additional
curvature perturbation which exceeds the observed value in the cosmic
microwave background (CMB) radiation~\cite{Ade:2015cva}.
Thus inflationary magnetogenesis
faces three obstacles: (i) substantial dilution of the produced magnetic fields after inflation, (ii) too large electromagnetic energy spoiling inflation itself, and (iii) induced curvature perturbations being inconsistent with the CMB observation.

In this paper, to overcome these three obstacles, we consider the axial
coupling model in which a (pseudo-)scalar field $\phi$ is coupled to the
electromagnetism through a term, $\phi F^{\mu\nu}\tilde{F}_{\mu\nu}$,
where $F_{\mu\nu}$ is the field strength tensor of the $U(1)$ gauge field, and
$\tilde{F}_{\mu\nu}$ is its dual. This coupling naturally arises under symmetries as the leading interaction of a pseudo-scalar and excites either of the left- or right-handed polarization modes of the gauge field. 
Helical magnetic fields, which are characterized by an imbalance between their two polarization states, have an interesting feature that part of its energy on
small scales is transferred to that on large scales by non-linear effects of
turbulent plasma with a high electric conductivity.
This phenomenon is called the {\it inverse cascade
process}. In practice, this process amplifies the
comoving amplitudes of helical magnetic fields on large scales, and,
as a result, their characteristic scale can grow.
Therefore it can alleviate the substantial dilution of such magnetic fields on large scales in the cosmological context~\cite{Banerjee:2004df, Durrer:2013pga}.
Recently, the possible signatures of cosmological helical
magnetic fields are reported in ref.~\cite{Tashiro:2013ita,Chen:2014qva}.

In the previous works on the axial coupling models, only the slow-roll
regime of $\phi$ was considered~\cite{Garretson:1992vt,Anber:2006xt,
Durrer:2010mq,Caprini:2014mja}. 
To evaluate the generated magnetic fields,
they applied an approximated analytic solution of the gauge field 
which is valid only 
during the slow roll regime
without its backreaction onto the inflationary dynamics.
However, we find that the generation of
the electromagnetic fields is much more efficient at the end of the
slow-roll regime and during the subsequent oscillation phase of $\phi$.
Since the slow-roll approximation is no longer
valid in these regimes, we numerically solve the equations of
motion of these fields, consistently taking into account the backreaction
from the gauge field to the dynamics of $\phi$ and the background expansion.
As a result, much stronger magnetic fields can be obtained than the ones in the preceding studies with the same setup~\cite{Garretson:1992vt,Anber:2006xt}.

In our scenario, the electromagnetic fields are mainly generated right
after the end of inflation, and hence we do not need to worry about the
electromagnetic energy density spoiling inflation.  Furthermore, since the
peak scales of the produced magnetic fields are far smaller than the CMB
scales, they are free from the CMB constraints.
Therefore our scenario overcomes the three obstacles of inflationary
magnetogenesis, and
relatively strong magnetic fields can be generated.
Unfortunately, however, in the case where $\phi$ is the inflaton
in large-field models,
the strength of the resultant magnetic fields does not reach the
observational bound, eq.~\eqref{obs lower bound}, by a few orders of
magnitude.  


Two additional issues concern us in the scenario. One is the effect of electrically charged particles. We investigate the generation of magnetic fields by solving the coupled system of the background inflaton and the gauge field.
However, if a sufficient amount of charged particles are present at the time of the magnetic field production, they may change the dynamics of the gauge field in a non-trivial way~\cite{Bassett:2000aw}. To avoid such cumbersome complexity, we require the inflaton decay rate is suppressed such that the charged particles are negligible during the production period.
Nonetheless, since the reheating is delayed,
the dilution of the produced magnetic field before the onset of the inverse cascade process is inevitably significant.
The other issue is the non-linear effect of the axial coupling on the dynamics of the inflaton and of the gauge field.
With this non-linear interaction taken into account, the gauge field produced by the background inflaton 
inverse-decays to the inflaton's fluctuations, which then in turn produces the gauge field.
For the large value of the axial coupling constant, this successive non-linear process would change the dynamics of the inflaton and gauge field as shown
in ref.~\cite{Adshead:2015pva}.
Since fully handling the non-linear dynamics requires lattice simulations,
we only consider the coupling constant small enough to ignore the effect of the inflaton perturbation on the gauge field dynamics. 


The rest of this paper is organized as follows. 
In sec.~\ref{Model}, we introduce our setup and review the approximated solution for the gauge field in the slow-roll regime. In sec.~\ref{Numerical Result}, we show the result of our numerical calculation and describe the dynamics of the model. In sec.~\ref{Inverse Cascade Process}, 
we take into account the inverse cascade process as the conservation of the helicity of the produced magnetic fields. Following the evolution after reheating, we exhibit the computation of the present values of the field strength and the correlation length for a given coupling constant in the model.
In sec.~\ref{Reheating}, we discuss the effects of charged particles on reheating in various cases. In sec.~\ref{The inlfaton perturbation}, 
the amount of the inflaton perturbation generated by the inverse-decay of the gauge quanta is evaluated,
and the strongest reliable value of the effective magnetic field strength in our frame work is presented. Finally we conclude in sec.~\ref{Conclusion}.

\section{Model}
\label{Model}

In this paper we consider the axial coupling model with the following Lagrangian:
\begin{equation}
\mathcal{L}=\frac{1}{2}\partial_\mu \phi \partial^\mu \phi -V(\phi) -\frac{1}{4}F_{\mu\nu}F^{\mu\nu}-\frac{\alpha}{4f}\phi F_{\mu\nu}\tilde{F}^{\mu\nu},
\end{equation}
where $\phi$ is the inflaton,%
$F_{\mu\nu}\equiv \partial_\mu A_\nu - \partial_\nu A_\mu$ is the field strength tensor of the $U(1)$ gauge field $A_\mu$, and $\tilde{F}^{\mu\nu} \equiv \epsilon^{\mu\nu\rho\sigma}F_{\rho\sigma}/\left( 2\sqrt{-g} \right)$ is its dual.
Inspired by the axion case in which $\phi$ is a pseudo-Nambu-Goldstone boson coupled to the axial vector current of charged fermions and the $\phi F\tilde{F}$ coupling arises from the chiral anomaly, we introduce the coupling constant as $\alpha/f$ where $\alpha$ is a dimensionless constant and $f$ corresponds to the axion decay constant. 
However, since this coupling is realized simply by symmetry consideration, we concentrate on phenomenology of the above Lagrangian 
without assuming any particular particle physics model.
For simplicity,  the potential is assumed to be quadratic form,%
\footnote{
The recent Planck result \cite{Ade:2015lrj} disfavors the quadratic potential of inflaton. In this work, however, we use it as an example to depict the consistent generation and evolution of the magnetic fields from the $\phi F \tilde F$ term. Moreover, as we will show later, the dominant contribution to the magnetic fields in this model comes from the modes that cross the horizon near the end of inflation, and their amplification due to parametric resonance occurs during the early stage of inflaton oscillation. During this period, the quadratic potential is expected to be a good approximation of less simple forms, justifying its use. 
}
\begin{equation}
V(\phi) = \frac{1}{2}m^2 \phi^2,
\quad
m=1.9\times10^{13}\GeV.
\label{inflaton-pot}
\end{equation}
In the flat FRW universe with the Einstein gravity, one can derive the equations of motion for the background inflaton $\phi_0(t)$ and the mode function of the gauge field $\mcA_\pm(k,t)$ as~\cite{Barnaby:2011vw}
\begin{align}
&\ddot{\phi_0}(t)+3H\dot{\phi}_0(t) +m^{2}\phi_0(t)
= \frac{\alpha}{f}\langle \bm{E}\cdot\bmB\rangle,
\label{EoM inflaton sim}
\\
& \ddot{\mcA}_{\pm}(k,t) +H \dot{\mcA_\pm}(k,t) +
\left(\frac{k^2}{a^2} \mp \frac{\alpha}{f} \frac{k}{a} \dot{\phi}_0 \right)
\mcA_\pm(k,t)=0,
\label{EoM gauge sim}
\end{align}
with
\begin{equation}
\langle \bm{E}\cdot\bmB\rangle=-\frac{1}{a^4} \int \frac{\dd^3 k}{(2\pi)^3}
\frac{k}{2} \, a \, \frac{\dd}{\dd t} \big[ |\mcA_+|^2 -|\mcA_-|^2 \big],
\label{EB sim}
\end{equation}
where the overdot denotes the cosmic time derivative, and $a$ is the scale factor.
In this paper, we take the Coulomb gauge, $A_0 = \partial_i A_i = 0$.
We decompose and quantize the gauge field as
\begin{equation}
 A_i(t, \bm{x})
 = 
 \sum_{\lambda=\pm} \int \frac{{\rm d}^3 k}{(2\pi)^3} 
 e^{i \bm{k \cdot x}} e_{i}^{(\lambda)}(\hat{\bm{k}}) 
 \left[ a_{\bm{k}}^{(\lambda)} \mcA_\lambda(k,t) 
  + a_{-\bm{k}}^{(\lambda) \dag} \mcA_\lambda^*(k,t) \right]
\,,
\label{quantization}
\end{equation}
where $e^{(\pm)}_i(\hat{\bm{k}})$ are the right/left-handed polarization vectors which satisfy $\epsilon_{ijl} k_j e_l^{(\pm)}(\hat{\bm{k}})=\mp i k e_i^{(\pm)}(\hat{\bm{k}})$,
and 
$a_{\bm{k}}^{(\pm) \dag}, a_{\bm{k}}^{(\pm)}$
are the creation/annihilation operators which satisfy the usual commutation relation, $[a^{(\lambda)}_{\bm{k}},a^{(\sigma) \dag}_{-\bm{k}'}]
= (2\pi)^3\delta(\bm{k}+\bm{k}')\delta^{\lambda \sigma}$.
With the definition of the electric and magnetic fields in terms of the gauge field, $E_i\equiv - a^{-1}\partial_t A_i$ and $B_i \equiv a^{-2} \epsilon_{ijk}\partial_j A_k$, one can derive eq.~\eqref{EB sim}.

Eq.~\eqref{EoM gauge sim} implies that either of two polarization modes $\mcA_\pm$ has the tachyonic instability for $k/a < \alpha |\dot{\phi}_0|/f$, depending on the sign of $\dot{\phi}_0$. Thus the gauge fields are produced
at the expense of the time kinetic energy of $\phi_0$. The term on the right hand side in eq.~\eqref{EoM inflaton sim} represents the backreaction from the
gauge field to the inflaton and it slows down the motion of $\phi_0$. 
It should be noted that the perturbation of the inflaton, $\delta\phi(t,\bm{x})$, is ignored in the above set of equations. We discuss the validity of this assumption in
sec.~\ref{The inlfaton perturbation}.

Before exploring our scenario with the full numerical calculation, it is instructive to review the gauge field production in the slow-roll regime of $\phi$. For this purpose,
it is useful to introduce a new parameter $\xi$ as
\begin{equation}
\xi\equiv \frac{\alpha \dot{\phi}_0}{2f H},
\end{equation}
where $H$ is the Hubble parameter. With the conformal time $\eta$,
eq.~\eqref{EoM gauge sim} is rewritten as
\begin{equation}
\left[ \partial_\eta^2 +k^2 \pm 2k \frac{\xi}{\eta} \right] \mcA_\pm(k,\eta)=0.
\end{equation}
When $\xi$ is constant which is a good approximation in the slow-roll regime, the exact solution of this equation is available.
If $\dot{\phi_{0}}<0$, then $\xi<0$ and the $(-)$ mode has the tachyonic instability.
With the adiabatic initial condition, 
\begin{equation}
\mcA_\pm (|k\eta|\gg1)=\frac{1}{\sqrt{2k}}e^{-ik\eta \mp i\xi\ln\vert k\eta \vert}, 
\label{initial condition}
\end{equation}
in the sub-horizon limit, one finds the analytic solution and the asymptotic expression for the growing $(-)$ mode are given by
%
\begin{equation}
\mcA_- (k,\eta) =
\frac{1}{\sqrt{2k}} \Big[ G_0 \left( \vert\xi\vert , - k \eta \right) + i F_0 \left( \vert\xi\vert , - k \eta \right) \Big]
\xrightarrow{|k\eta|\ll 1}
\frac{1}{\sqrt{2k}} \frac{e^{\pi|\xi|}}{\sqrt{2\pi|\xi|}},
\end{equation}
%
%
up to an irrelevant constant phase, where $F_L(\kappa,z)$ and $G_L(\kappa,z)$ are regular and irregular Coulomb wave functions, respectively.
Therefore the growing mode is amplified by a factor $e^{\pi|\xi|}/\sqrt{2\pi|\xi|}$
due to the tachyonic instability. Since the produced gauge field induces non-gaussian perturbations of the inflaton, $\delta\phi$, the value of $\xi$
at the horizon-crossing of the CMB mode ($k_*=0.002\Mpc^{-1}$)
is constrained as $|\xi_*|<2.37$~\cite{Pajer:2013fsa}.
At smaller scales, non-detection of primordial black holes (PBHs) can potentially push the constraint further down \cite{Linde:2012bt}. While the constraint from PBHs inherits uncertainty, the forthcoming generations of terrestrial gravitational-wave (GW) detectors can directly probe the GWs sourced by the produced gauge field. It has been demonstrated that, for a quadratic potential of inflaton $\phi$, the second-generation detectors (Advanced LIGO, VIRGO, and KAGRA) will be able to probe $\vert\xi_*\vert \gtrsim 2.2$ (at $95\%$ CL, based on flat priors on $\xi$) and the third-generation experiments can improve the sensitivity to $\vert\xi_*\vert \gtrsim 1.9$ \cite{Barnaby:2011qe,Crowder:2012ik}.

\begin{figure}[tbp]
 \begin{center}
  \includegraphics[width=75mm]{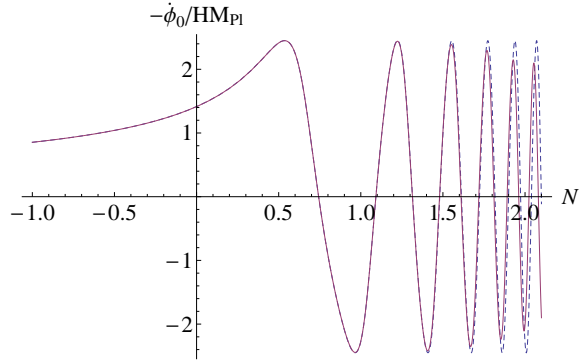}
 \end{center}
 \caption
 { The time evolution of $\dot{\phi}_0/H \propto \xi$ is shown. 
 The horizontal axis denotes the e-folding number $N$, and
 inflation ends $(\epsilon_H \equiv-\dot{H}/H^2 \ = 1)$ at $N=0$.
The initial conditions are $\phi_0=\sqrt{282}\Mpl$ and $\dot{\phi}_0=-\sqrt{2/3}\Mpl$
without any gauge fields.
The blue dashed line shows the case with $\alpha=0$, while the red line shows the case with $\alpha\Mpl/f=8$. 
In both cases, $\dot{\phi}_0/H$ (or $\xi$) increases during inflation and reaches its maximum value after the end of inflation. Therefore the most efficient production
of the gauge fields takes place after inflation.
}
 \label{phidot}
\end{figure}

Since the gauge field production is characterized by $|\xi|$, it is expected that the generation
of the gauge fields is the most efficient when $|\xi|$ reaches its maximum value. Indeed, $\xi\propto \dot{\phi}_0/H \simeq -2\Mpl^2/\phi_0$ in the slow-roll regime, and its absolute value increases during inflation.
In fig.~\ref{phidot}, we plot the evolution of $\dot{\phi}_0/H\propto \xi$.
%
%
One can see that $|\xi|$ increases during inflation and reaches its maximum value when $\phi_0$ crosses its origin at $N\approx 0.5$ after inflation. It then starts to oscillate with a large
amplitude in the oscillation phase. Thus, in order to evaluate the produced magnetic field, one has to investigate the dynamics after inflation, the period during which the analytic solution is no longer valid.
It seems that this critical point is overlooked in the previous works~\cite{Garretson:1992vt,Anber:2006xt,
Durrer:2010mq,Caprini:2014mja}.
Note that the amplitude of $\xi$ slightly decreases and the oscillation phase changes for $N\gtrsim1.5$ in the case with $\alpha\Mpl/f=8,$ because of the backreaction from the gauge field (compare the two lines in fig.~\ref{phidot}).

\section{Numerical Result}
\label{Numerical Result}

By a numerical calculation, we simultaneously solve eqs.~\eqref{EoM inflaton sim}, \eqref{EoM gauge sim}
and the Friedmann equation,
\begin{equation}
3\Mpl^2 H^2(t) = \frac{1}{2} \left[\dot{\phi_0}^2(t) +m^2 \phi_0^2(t)\right] +\rho_\em(t),
\label{Friedmann eq sim}
\end{equation}
where $\Mpl$ is the reduced Planck mass, and the energy density of electromagnetic fields $\rho_\em$ is given in terms of the mode function of the gauge field $\mcA_\pm(k,t)$ as
\begin{align}
&\rho_\em =\rho_\em^{(-)}+ \rho_\em^{(+)},\qquad
\rho_\em^{(\pm)}=\frac{1}{2} \int \frac{\dd k}{k} \big[\mcP_E^{(\pm)} + \mcP_B^{(\pm)}\big],
\label{total rho sim}
\\
&\mcP_E^{(\pm)} = \frac{k^3  |\dot{\mcA}_{\pm}|^2}{2\pi^2a^2},
\qquad
\mcP_B^{(\pm)} = \frac{k^5  |\mcA_{\pm}|^2}{2\pi^2a^4}.
\label{EB power spectra}
\end{align}
However, since the gauge field is solved in Fourier space, and $\langle \bm{E}\cdot\bmB\rangle$ and $\rho_\em$ contain the momentum integrations,
we have to solve eq.~\eqref{EoM gauge sim} for many modes of $\mcA_\pm(k,t)$ with different wave numbers simultaneously.
In our numerical calculation, we start to solve for a mode function $\mcA_\pm(k,t)$ with the initial condition eq.~\eqref{initial condition}, when its wave number equals to a rapidly growing function, $100k_{\rm ins}(t)$, where $k_{\rm ins} \equiv \alpha a \dot{\phi}/f =2\xi aH$ is the wave number for which the $(-)$ mode becomes unstable, and continue to solve the mode until the end of calculation. In other words, we solve for all modes which satisfy
$100k_{\rm ins}(t_i)<k< 100k_{\rm ins}^{\rm max} (t)$, where $t_i$ is
the onset of the calculation, $N(t_{i})\approx-70$, and $k_{\rm
ins}^{\rm max} (t)$ is the maximum value of $k_{\rm ins}$ on the
interval $[t_i,t]$. The grid size of the wave number is set as $\delta \ln k = 0.01.$
To eliminate the contribution from the modes in the vacuum state, we do not integrate all the modes being solved in eqs.~\eqref{EB sim} and \eqref{total rho sim} but only the modes satisfying $k<k_{\rm ins} (t)$ at least once during the numerical evolution. 
For confirmation, we check another criterion. We integrate only the modes with
a sufficient amplitude,
$|\sqrt{2k}\mcA_\pm| > 20$,
and compare the result with that of the former criterion, confirming no
recognizable change.
The time step of the numerical calculation is set as  
$\delta t = 10^{-4} a(t)/k_{\rm ins}(t)$,
which is always much shorter than the inflaton oscillation time scale, $m^{-1}$.
We check the total energy conservation, $\dot{\rho}=-3H(\rho+p)$, is satisfied in sub-percent accuracy.

In this section, we discuss only the case of $\alpha\Mpl/f = 8$, because
it is close to the maximum value of the coupling constant $\alpha\Mpl/f$ for which our treatment is justified.
For larger $\alpha\Mpl/f$, we need to take into account the inflaton perturbation
induced by the gauge field. 
However, it requires lattice simulations as the re-scattering process between the gauge field $\mcA_\pm$ and $\delta\phi$ is fully non-linear. 
Thus in this paper, we only consider the case where the coupling is not too strong and the inflaton perturbation is negligible, $\delta\phi\ll\phi_0$.
We will discuss this point with more detail in sec~\ref{The inlfaton perturbation}.

Let us now show the result of our numerical calculation.
%
\begin{figure}[tbp]
  \hspace{-2mm}
  \includegraphics[width=75mm]{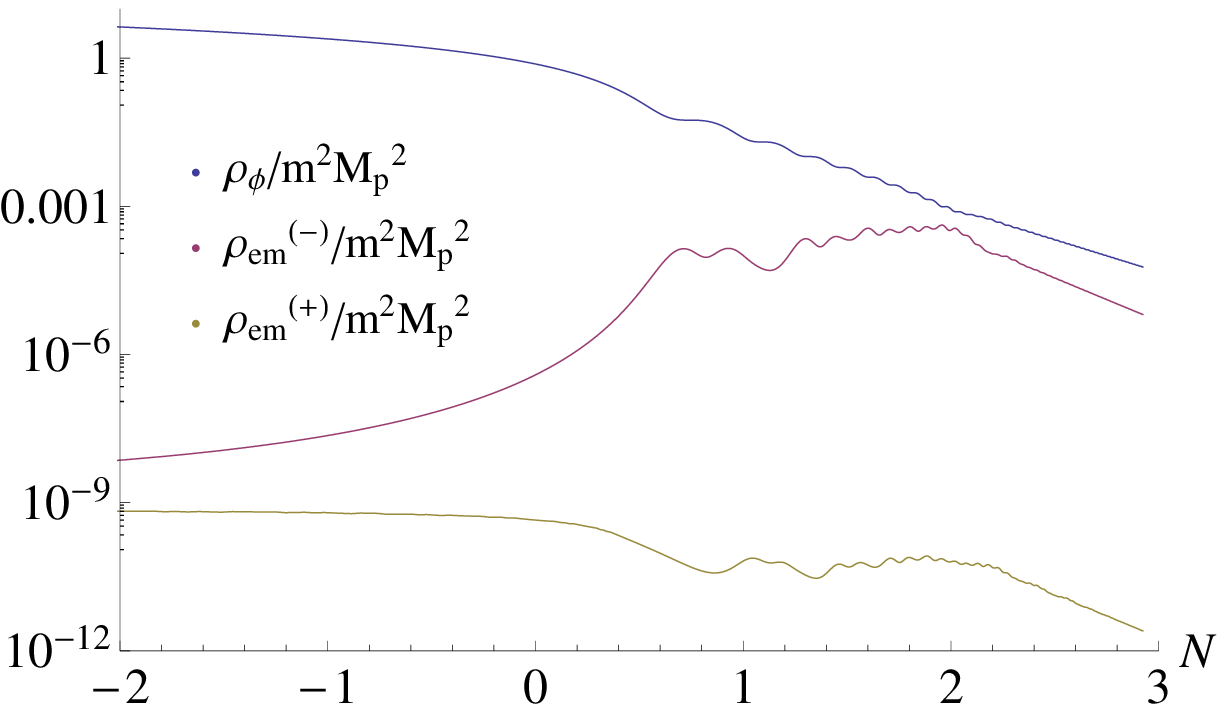}
  \hspace{5mm}
  \includegraphics[width=75mm]{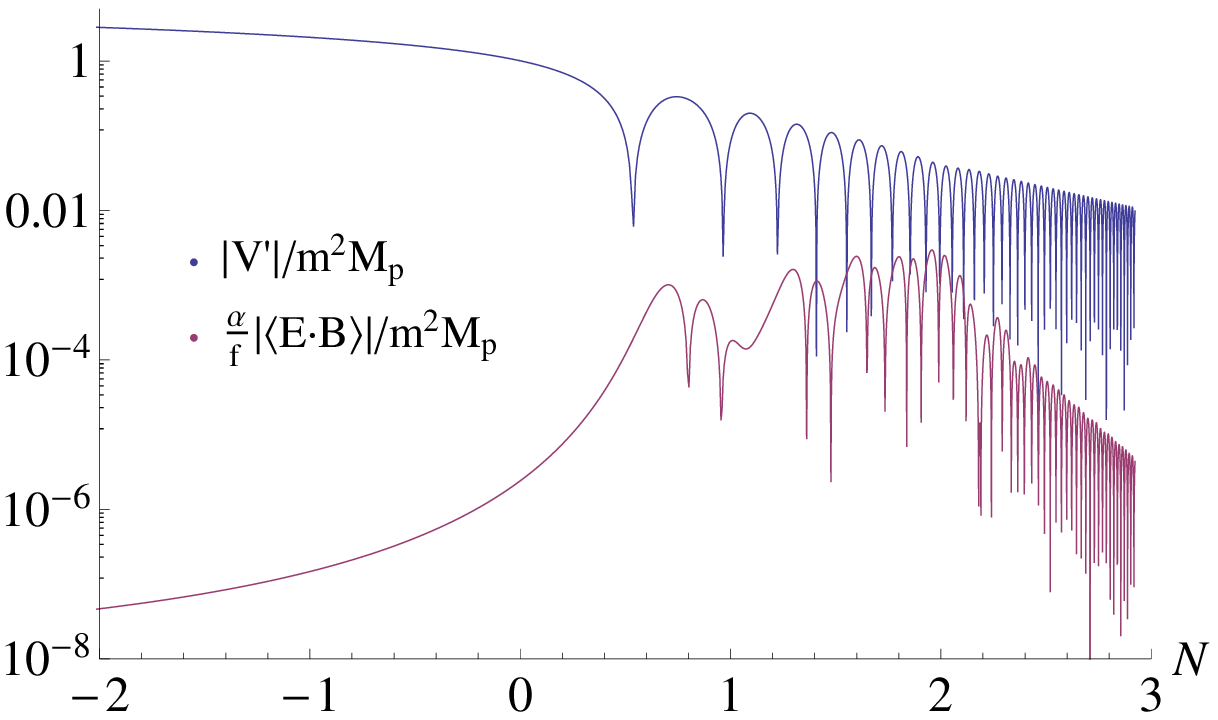}
  \caption
 { The energy density of the inflaton $\rho_\phi$ (blue line), the $(-)$ mode of the gauge field $\rho^{(-)}$ (red line) and the $(+)$ mode of the gauge field $\rho^{(+)}$ (yellow line) are shown in the left panel. The potential force $V' \equiv \partial V / \partial\phi=m^2\phi$ (blue line) and the backreaction term $\alpha \langle\bm{E}\cdot\bm{B}\rangle/f$
(red line) in eq.~\eqref{EoM inflaton sim} are compared in the right panel.
We take $\alpha \Mpl/f =8$.
The tachyonic instability and the subsequent parametric resonance
amplify the $(-)$ mode while the $(+)$ mode is not significantly produced.
The energy densities eventually decay as $\rho_\em \propto a^{-4}$ and $\rho_\phi\propto a^{-3}$ after the coupling between $\phi$ and $\mcA_\pm$ becomes ineffective at $N\approx 2.3$.}
\label{rho_evolution}
\end{figure}
In fig.~\ref{rho_evolution}, we plot the energy density of the inflaton and
the two polarization mode of the gauge field in the left panel. The horizontal axis $N$ denotes the e-fold number, and inflation ends at $N=0$.
One can clearly see that the electromagnetic fields are mainly generated
after inflation and the energy density $\rho_\em = \rho_\em^{(-)}+\rho_\em^{(+)}$ reaches its maximum value at 2 e-folds after
the end of inflation. Note that $\rho_\em$ is always dominated by the contribution
from the $(-)$ mode $\rho_\em^{(-)}$. 
Until $N\simeq 0.75$, $\dot{\phi}$ is negative (see fig.~\ref{phidot}), and only the $(-)$ mode of the gauge field continues to grow due to the tachyonic instability. 
However, after the sign of $\dot{\phi}$ changes, the $(-)$ mode transiently decreases due to acquiring a positive mass, while the $(+)$ mode begins to increase by the tachyonic instability in turn. 
Subsequently, the inflaton oscillation develops, and the parametric resonance becomes efficient. 
Although not only $\mcA_-$ but also $\mcA_+$ is amplified in this regime, $\mcA_-$ is far larger than $\mcA_+$ due to the initial hierarchy implemented by the tachyonic instability, and thus a helical magnetic field is generated.
As the amplitude of the inflaton oscillation decreases, the efficient resonance eventually terminates. After that, since the inflaton and the gauge field are now decoupled, they behave as the usual matter and radiation components, respectively.

In the right panel of fig.~\ref{rho_evolution}, the potential force of the inflaton $V' \equiv \partial V / \partial\phi=m^2\phi$ and the backreaction from the gauge field to the inflaton $\alpha \langle\bm{E}\cdot\bm{B}\rangle/f$ are shown.
Although the backreaction term is always subdominant in this case of $\alpha\Mpl/f = 8$, it becomes non-negligible and the behavior of $\phi$ is significantly affected for a larger coupling constant.

%
\begin{figure}[tbp]
  \hspace{-2mm}
  \includegraphics[width=75mm]{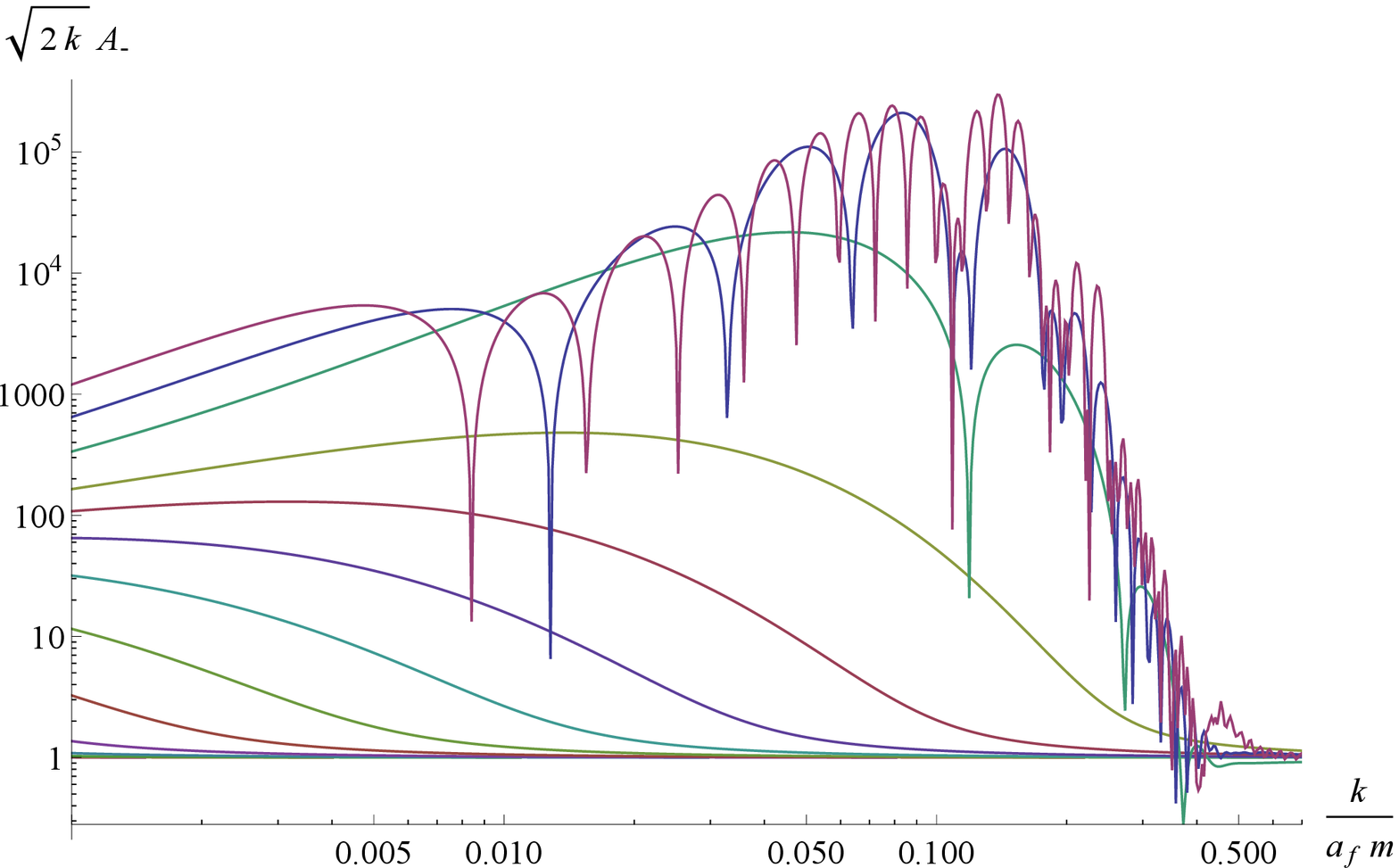}
  \hspace{5mm}
  \includegraphics[width=75mm]{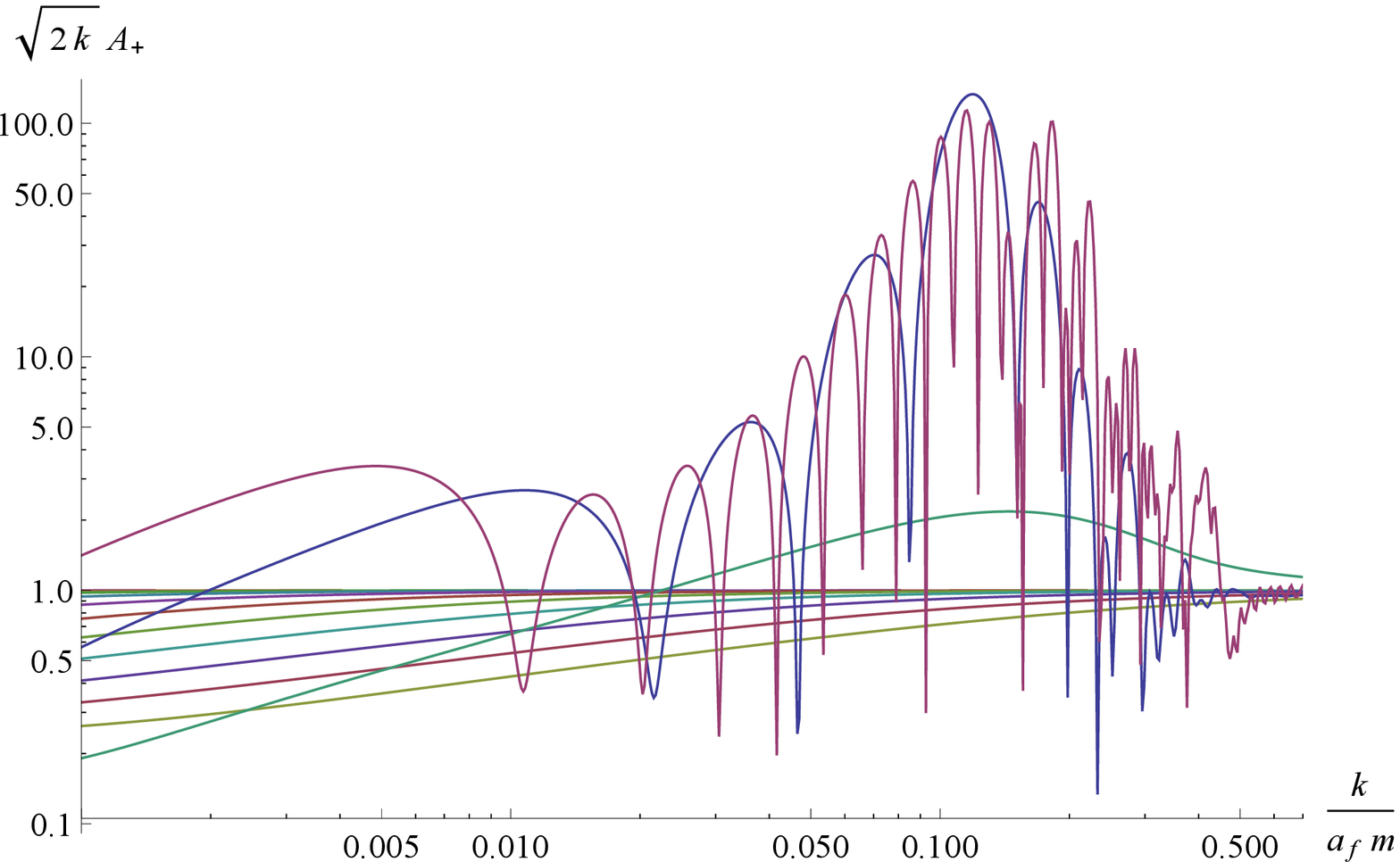}
  \caption
 { 
The time evolution of the spectra of the mode function $\mcA_\pm(k,t)$.
 The horizontal axis denotes the physical wave number at the end of the calculation, $N_f \approx 2.9$.
Colored lines represent the spectra at different times with time steps of $1$ e-fold from bottom to top, and the top red lines denote the spectra at $N_f$. 
$\mcA_-$ (left panel) acquires the double amplifications both from the tachyonic instability and the parametric resonance, and the peak mode reaches amplitudes of $\mathcal{O}(10^5)$.
$\mcA_+$ (right panel) is slightly damped during inflation due to the positive effective mass, while the parametric resonance amplifies the peak mode by $\mathcal{O}(10^2)$.}
 \label{Apm_spectrum}
\end{figure}
%
In fig.~\ref{Apm_spectrum}, we show the spectra of the mode function $\mcA_\pm (k,t)$. 
Although $\mcA_+$ does not have the tachyonic instability during inflation,
the oscillating inflaton amplifies $\mcA_+$ to at most  $\mathcal{O}(10^2)$ through the parametric resonance as the oscillating behavior of the spectrum indicates.
The mode~$\mcA_-$ acquires the double amplification both from the tachyonic instability and the parametric resonance, and the peak mode at $k/a_f \sim 0.1 m$ undergoes the $\mathcal{O}(10^{5})$ amplification.
It corresponds to $|\xi|\approx 4.5$ in the constant $\xi$ case,
while $\xi$ at the horizon crossing of the CMB scale
is $\vert\xi\vert\simeq \alpha\Mpl/2\sqrt{4N_\CMB+2}f \approx 0.3$ in this setup.
It should be noted that since $\mcA_-$ is much larger than $\mcA_+$ at the end, an almost maximal helical magnetic field is eventually generated (see eq.~\eqref{def helicity}).

%
\begin{figure}[tbp]
  \hspace{-2mm}
  \includegraphics[width=75mm]{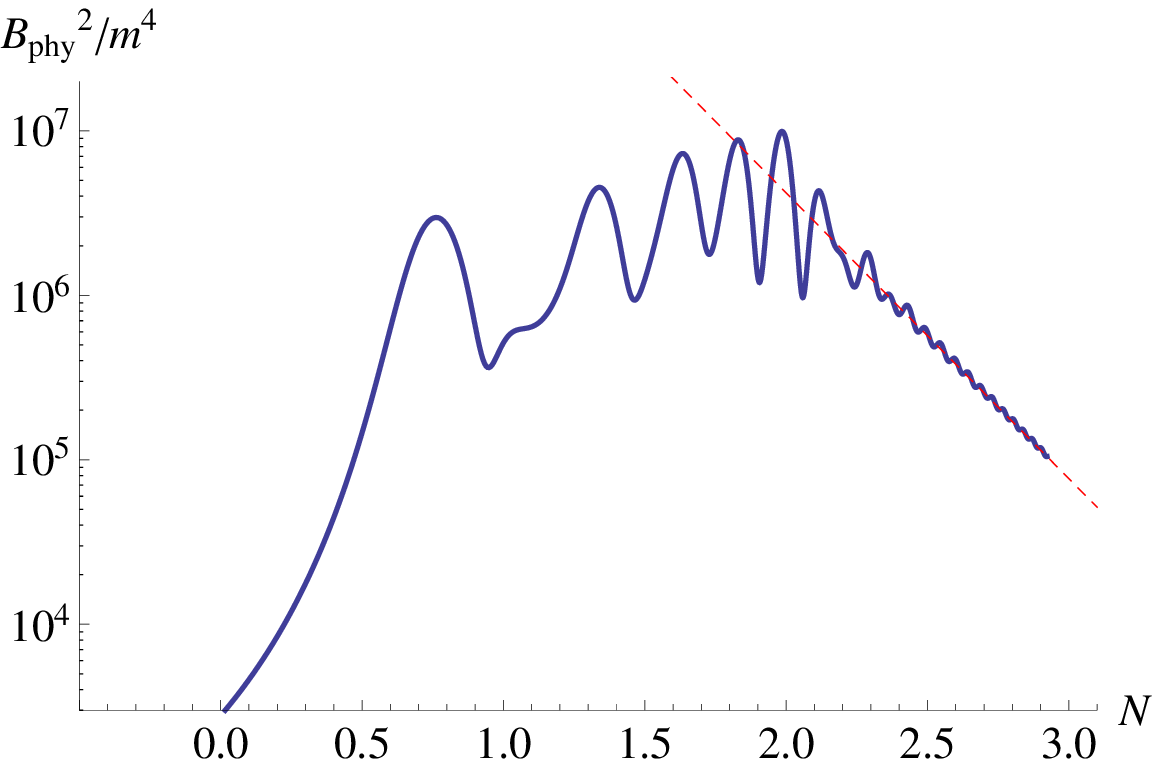}
  \hspace{5mm}
  \includegraphics[width=75mm]{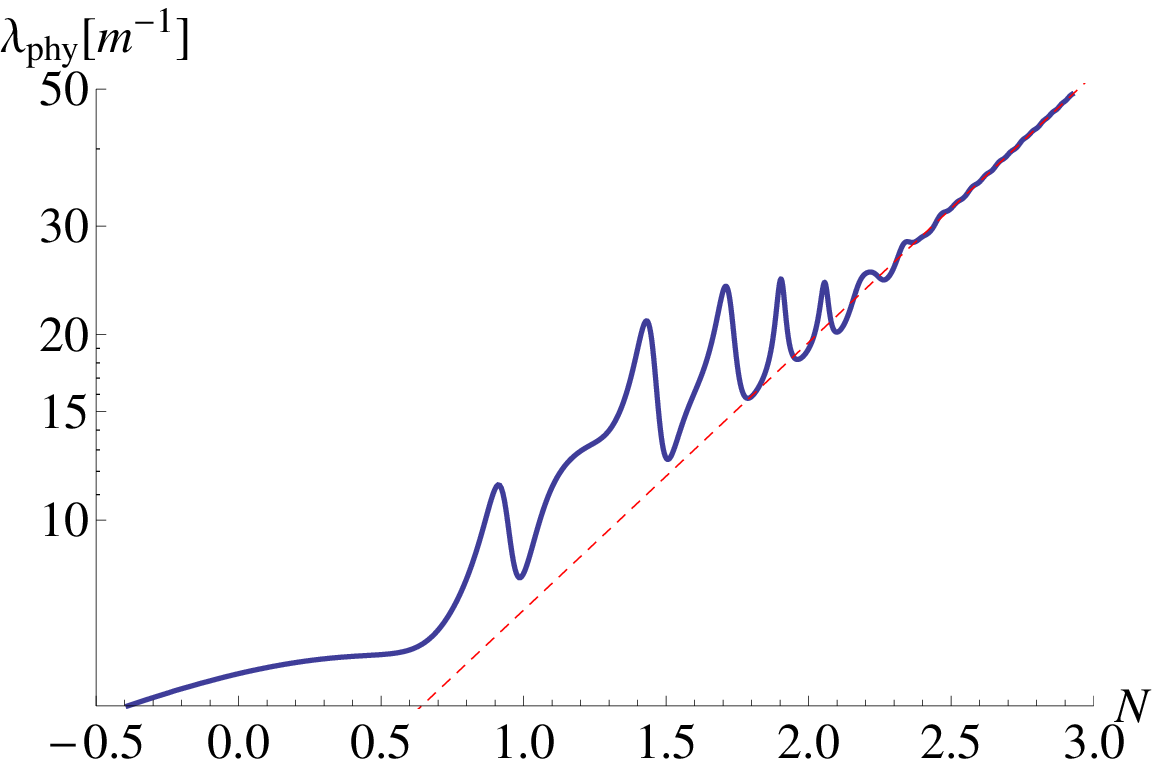}
  \caption
 { The time evolution of the physical intensity $B_{\rm phy}^2$ (left panel)
 and the physical correlation length (right panel) of the magnetic field are shown. 
The e-fold number $N$ is normalized such that inflation ends at $N=0$.
After the decoupling ($N\gtrsim2$), they behaves as 
$B^2_{\rm phy} \propto a^{-4}$, and
$\lambda_{\rm phy}\propto a$ (red dashed lines). 
However, this behavior will be changed when charged particles appear and begin to interact with the electromagnetic fields. }
 \label{B_spectrum}
\end{figure}
%
Following refs.~\cite{Durrer:2013pga, Caprini:2014mja}, we define the physical intensity and the physical correlation length of the magnetic field as
\begin{equation}
B^2_{\rm phy}(t)\equiv a^{-4} \int \frac{\dd^3 k}{(2\pi)^3}\, k^2 \Big(
|\mcA_+|^2+|\mcA_-|^2\Big),\quad
\lambda_{\rm phy}(t) \equiv a \frac{\int \dd^3 k\,\frac{2\pi}{k}\, k^2 \left(
|\mcA_+|^2+|\mcA_-|^2\right)}{\int \dd^3 k\, k^2 \left(
|\mcA_+|^2+|\mcA_-|^2\right)}.
\end{equation}
In fig.~\ref{B_spectrum}, we plot them as a function of the e-fold number $N$ for $\alpha \Mpl/f =8$.
After the coupling between the inflaton and the gauge field becomes ineffective
($N\gtrsim 2$), the electromagnetic fields effectively satisfy the free wave equation,  $[ \partial_\eta^2 +k^2]\mcA_\pm(k,\eta)=0$, and thus $|\mcA_\pm|$ become constant. Then we can fit the numerical result as
(see the red dashed lines in fig.~\ref{B_spectrum}.)
\begin{equation}
B^2_{\rm phy}\simeq 1.3\times 10^{10} m^4 e^{-4N},
\quad
\lambda_{\rm phy}\simeq 2.6m^{-1} e^{N},
\quad (N\gtrsim2).
\label{Blambda N}
\end{equation}
By using the fitted behavior of the Hubble parameter, $H\simeq 0.36m e^{-3N/2}$,
and the unit conversions $1\G=6.8\times10^{-20}\GeV^2$ and $1\GeV^{-1}=6.4\times10^{-39}\Mpc$, one can rewrite eq.~\eqref{Blambda N} as
\begin{equation}
B^2_{\rm phy}\simeq 5.4\times 10^{102} \left(\frac{H}{m}\right)^{8/3}  \G^2,
\quad
\lambda_{\rm phy}\simeq 4.5\times 10^{-52}\left(\frac{H}{m}\right)^{-2/3} \Mpc,
\quad (H\lesssim 10^{-2}m).
\label{produced MF}
\end{equation}
This is the physical magnetic field amplitude with the associated correlation length after all the production processes complete, and this behavior continues to hold until charged particles are generated by the inflaton decay.
Thereafter the produced helical magnetic field undergoes the inverse cascade process.

\section{Inverse Cascade Process}
\label{Inverse Cascade Process}

As the inflaton decays and particles in the standard model appear, the electric
conductivity $\sigma_c$ increases. The conductivity gives a friction
term in the equation of $\mcA_\pm$ as $[ \partial_t^2
+(H+\sigma_c)\partial_t +k^2/a^2]\mcA_\pm=0$. Since, after the reheating
completes, the conductivity of the thermalized plasma
exceeds the Hubble parameter, $\sigma_c \gg H$,
one can find that on large scales $k \ll a \sqrt{\sigma_c H}$
the gauge mode functions freeze, $\mcA_\pm\simeq const.$~\cite{Martin:2007ue}.
Thus the corresponding electric fields vanish, and the static magnetic fields
which merely dilute due to the cosmic expansion remain on such large scales.
On small scales, the evolution of magnetic fields is different
due to the nonlinearity of the fluid dynamics~\cite{Banerjee:2004df, Durrer:2013pga}.
Turbulence in a plasma induced by the Lorentz force highly develops on small scales. 
The turbulence transfers the energy of the magnetic fields from large scales
to small scales due to a non-linear interaction
between the plasma fluid and the magnetic fields.
This is called the {\it direct cascade} process. 
The resultant spectrum of the magnetic field is red-tilted on
small scales where the time scale of the turbulence is smaller than
the cosmological time scale.

However, when a magnetic field is helical, not all of its energy is transferred to small scales.
According to Maxwell equations, the magnetic helicity should be conserved in the magnetized plasma with a high
conductivity. Therefore, in order to maintain the helicity,
a part of the energy is transferred to a slightly larger scales where the
turbulence does not develop well. This is called the {\it inverse cascade} process.

The magnetic helicity $\mcH$ is defined as
$\mcH\equiv \int_V \dd^3 x \epsilon_{ijk}  A_i \partial_j A_k$
where the spacial integral is taken over a finite comoving volume $V$ on which 
magnetic fields have no component normal to the boundary, i.e. $\bmB\cdot \bm{n}=0$. Then $\mcH$ is invariant under the gauge transformation (see e.g.~\cite{Brandenburg:2004jv}).
With eq.~\eqref{quantization}, the expectation value of the helicity density $\mathfrak{h}$ which is defined as $\mathfrak{h}\equiv \mcH/V$ is given by
\begin{equation}
\langle \mathfrak{h}\rangle = 
\frac{1}{V}\int_V\dd^3 x\, \epsilon_{ijk} \langle A_i \partial_j A_k \rangle
= \int \frac{\dd^3 k}{(2\pi)^3}\, k \Big( |\mcA_-|^2 - |\mcA_+|^2\Big).
\label{def helicity}
\end{equation}
Thus the helicity density represents the difference between the
left-handed and the right-handed polarization modes. 
When a magnetic field is perfectly helical~(i.e.~one component of
polarization modes is non-zero while the other components are zero), 
the helicity conservation, $\mcH=const.$ 
or equivalently $\mathfrak{h}=const.$,
provides the constraint between
the strength and the correlation length of the helical magnetic field as
\begin{equation}
\langle\mathfrak{h}\rangle \propto a^3 \lambda_{\rm phy} B_{\rm phy}^2 =const. 
\label{Helicity conservation}
\end{equation}
It is clear that this equation holds for freezed magnetic fields because their physical scale and
strength simply evolved as $\lambda_{\rm phy}(t) \propto a$ and $B_{\rm phy}(t) \propto a^{-2}$. 
However this equation is valid even in the turbulent regime where 
the magnetic field evolution is affected by the turbulence in a plasma
and $\lambda_{\rm phy}$ grows faster than $a$.
The scale and strength of magnetic fields in the turbulent regime evolve with satisfying this equation. 

The study about the evolution of magnetic fields with non-zero helicity after inflation requires a MHD numerical simulation.
However the helicity conservation enable us to directly compare
helical magnetic fields produced in early universe with the observation bound by
using eq.~\eqref{Helicity conservation}, because the constraint from
blazar observations, eq.~\eqref{obs lower bound}, is proportional to $\lambda_{\rm phy}B_{\rm phy}^2
\propto \langle \mathfrak{h}\rangle$ for magnetic fields with $\lambda < 1 ~{\rm Mpc}$.
The scale factor at the completion of reheating is given by
\begin{equation}
a_{\reh} \approx 5.5\times10^{-32} \sqrt{\frac{\Mpl}{\Gamma_\phi}},
\quad (H=\Gamma_\phi),
\end{equation}
where $\Gamma_\phi$ is the decay rate of the inflaton, and we set the scale
factor at present as unity. Therefore, when the helicity conservation is valid,
the effective strength
of the helical magnetic field at present can be written as
\begin{equation}
B_{\rm eff} \simeq 1.3 \times 10^{-47} 
B_{\rm phy}(t_\reh) 
\left(\frac{\lambda_{\rm phy}(t_\reh)}{1\Mpc} \right)^{1/2}  
\left(\frac{\Mpl}{\Gamma_\phi} \right)^{3/4},
\quad (\lambda < 1\Mpc),
\label{late time evolution}
\end{equation}
where $t_\reh$ denotes the time at the reheating completion.
Recall $B_{\rm eff}$ is defined in eq.~\eqref{obs lower bound}.
Substituting eq.~\eqref{produced MF} with $H=\Gamma_\phi$,
we finally obtain the produced $B_{\rm eff}$ at the present epoch
in the case of $\alpha\Mpl/f=8$ as
\begin{equation}
B_{\rm eff}= 6.6\times 10^{-20}\G \left(\frac{\Gamma_\phi}{10^6 \GeV}\right)^{1/4},
\qquad
\left( \frac{\alpha \Mpl}{f}=8\right).
\label{Beff result}
\end{equation}
In the next section, we discuss the allowed values of $\Gamma_\phi$. 
Furthermore, in sec.~\ref{The inlfaton perturbation}, we explore the larger values of the coupling constant $\alpha/f$. 
There exists the parameter space where the stronger $B_{\rm eff}$ is obtained.

Although, with the helicity conservation, we can estimate $B_{\rm eff}$
at present from the strength and scale of a helical magnetic field at the reheating epoch, 
$B_{\rm phy}(t)$ and $\lambda_{\rm phy}(t)$ are degenerate in $B_{\rm eff}$.
Before closing this section, we explore the way to resolve the degeneracy between
$B_{\rm phy}(t)$ and $\lambda_{\rm phy}(t)$.

If the initial spectrum of the helical magnetic fields is blue-tilted, as in the present case, its characteristic length scale, at which most of the magnetic field energy resides, is shifted to the scale comparable to that of the turbulence, $\lambda_T(t)$. This is due to the direct cascade process, in which 
the magnetic field energy on the scales smaller than $\lambda_T(t)$
is significantly transferred by the turbulence to farther smaller scales
and dissipates away.

The turbulence scale $\lambda_T$ is given by the plasma fluid velocity $v$
and the cosmic time $t$ as $\lambda_T(t) \sim v t$.
When the turbulence fully develops, the
equipartition state between the fluid and the magnetic field is achieved. In that
case, $v$ reaches the Alfv\'en velocity, $v_A(t) \simeq
B_{\rm phy}/\sqrt{\rho}$, where $\rho$ is the energy density of the
plasma.
Thus one finds
$\lambda_{\rm phy} \sim t B_{\rm phy}/\sqrt{\rho}.$ 
This equation holds until the recombination when
most of the charged particles disappears and the turbulence
terminates~(see also refs.~\cite{Banerjee:2004df,Durrer:2013pga} for a
more detailed discussion).
After that, $B_{\rm phy}$ and $\lambda_{\rm phy}$ evolve as usual, 
only due to the cosmological expansion, $B_{\rm phy}\propto a^{-2}, \lambda_{\rm phy} \propto a$.%
~\footnote{Since residual charged particles remain after the
recombination, the MHD interaction is effective between the magnetic fields and
the plasma fluid. However, since the Alfv\'en velocity is proportional
to $a^{-1/2}$ in the matter dominated plasma, $t v_A$ evolves as
$\propto a$. Therefore, the MHD interaction in the matter dominated era does not induce the
further evolution of $\lambda_{\rm phy}/a$. }
Thus, $B_{\rm phy}/\lambda_{\rm phy}$ at present is given by
%
\begin{align}
\frac{B_{\rm phy}}{\lambda_{\rm phy}}(t_{\rm now}) \sim \frac{\sqrt{\rho(t_{\rm rec})}}{t_{\rm rec}}a_{\rm rec}^3
\sim \frac{\rho(t_{\rm rec})}{\Mpl}a_{\rm rec}^3
\sim\frac{T_{\rm CMB}^3T_{\rm rec}}{\Mpl},
\end{align}
where the subscript ``rec'' denotes the recombination and $T_{\rm CMB} \approx 3$K.
Therefore, we obtain~\cite{Banerjee:2004df,Durrer:2013pga}
\begin{equation}
B_{\rm phy}(t_{\rm now}) \sim 10^{-8} \G \left(\frac{\lambda_{\rm phy}(t_{\rm now})}{1\Mpc}\right).
\label{Blambda relation 1}
\end{equation}
Combining this equation and the helicity conservation, one can solve for each of $B_{\rm phy}$ and $\lambda_{\rm phy}$ at present as
\begin{equation}
B_{\rm phy}(t_{\rm now}) \sim10^{-8}\mathcal{X}\G,\quad 
\lambda_{\rm phy}(t_{\rm now})\sim\mathcal{X} \Mpc,
\quad\mathcal{X}\equiv a_\reh\left(\frac{B_{\rm phy}(t_\reh)}{10^{-8}\G} \right)^{2/3} \left(\frac{\lambda_{\rm phy}(t_\reh)}{1\Mpc} \right)^{1/3}.
\end{equation}
Substituting eq.~\eqref{produced MF} with $H=\Gamma_\phi$, we find
\begin{equation}
\mathcal{X} \approx3.5\times10^{-8} 
 \left(\frac{\Gamma_\phi}{10^6 \GeV}\right)^{\frac{1}{6}},
 \qquad \left( \frac{\alpha \Mpl}{f}=8\right).
\end{equation}
The MHD numerical simulations in ref.~\cite{Banerjee:2004df,Christensson:2000sp,Kahniashvili:2012uj}
show that, due to the inverse cascade,
the spectral index of the magnetic field energy spectrum, ${\cal P}_B$,
is $\sim 5$ on the scales larger than the peak
scale.
Therefore, the magnetic field strength at the scale $L$~($L>\lambda_{\rm phy}$) is roughly evaluated as
$B_L \sim B_{\rm phy} (\lambda_{\rm phy}/L)^{5/2}$.

\section{Reheating}
\label{Reheating}

Provided that reheating proceeds instantaneously and 
charged particles appear at once, one can substitute eq.~\eqref{produced MF} with $H=\Gamma_\phi$ into eq.~\eqref{late time evolution} 
independently on the value of $\Gamma_\phi$, 
(if $\lambda<1\Mpc$ and $\Gamma_\reh\lesssim10^{-2}m$).
However, charged particles can be produced before the completion of reheating, $H=\Gamma_\phi$, since the inflaton decay is active even for $H>\Gamma_\phi$.
For example, by assuming the inflaton decay rate $\Gamma_\phi$ is constant and the energy density of the decay products $\rho_\rad(t)$ satisfies,
$\dot{\rho}_\rad +4H \rho_\rad=\Gamma_\phi \rho_\phi$,
one finds $\rho_\rad = 2\Gamma_{\phi} \rho_\phi/5H$ in the inflaton
oscillating phase~\cite{Kolb:1990vq}. Thus the decay products occupy roughly $\Gamma_\phi/H$ of the total energy density.
If the charged particles in $\rho_\rad$ affected the dynamics of $\mcA_\pm$
before the decoupling between the inflaton and the gauge field,
we would have to reconsider the result in sec.~\ref{Numerical Result},
which does not take into account the effect of the charged particles.%
\footnote{In ref.~\cite{Bassett:2000aw}, the authors
study how the growing electric conductivity suppresses the generation
of magnetic fields by a parametric resonance, while they assume
different couplings, $\phi^2A^2$ and $RA^2$, from ours.}
Therefore in this section, we discuss the maximum value of $\Gamma_\phi$ for which our numerical result in sec.~\ref{Numerical Result} is verified. 

\subsection{Case 1 : the inflaton decays into charged particles}

First, we consider the case where the inflaton mainly decays into 
charged particles with a 
constant decay rate $\Gamma_\phi$. 
The produced charged particles alter the dynamics
of the gauge field if the interaction between them is significant.
Since the cross section of the interaction is roughly $\sigma_{\rm int} \simeq \alpha'^2/p^2$
where $\alpha'$ is the fine structure constant, and $p$ 
denotes a typical momentum of the charged particle whose mass is
negligible compared to the inflaton mass $m$,
the interaction rate per a Hubble rate can be estimated as
\begin{equation}
\frac{\Gamma_{\rm int}}{H} =\frac{n_{\rm c} \sigma_{\rm int} v}{H}
\simeq \alpha'^2 \frac{\Gamma_\phi \Mpl^2}{m^3}
\approx \left(\frac{\alpha'}{0.01}\right)^2
\left(\frac{\Gamma_\phi}{10^7\GeV}\right) \left(\frac{m}{2\times10^{13}\GeV}\right)^{-3} ,
\label{case 1}
\end{equation}
where $n_{\rm c}\simeq \rho_\rad/m$ is the number density of the charged particles whose momentum is $p\simeq m$, and we have used $\rho_\rad\simeq \Gamma_\phi \rho_\phi/H \simeq \Gamma_\phi H \Mpl^2$. Therefore, for $\Gamma_\phi \ll 10^{7}\GeV$,
the charged particles do not significantly interact with the gauge field
within a Hubble time during the inflaton oscillating phase. Note that 
the electromagnetic fields are generated for an interval of 2 e-folds after the end of inflation (see fig.~\ref{rho_evolution}).
Thus $\Gamma_\phi$ does not have to be suppressed by many orders of magnitude
than $10^{7}\GeV$.

Eq.~\eqref{case 1} is also used as a condition of thermalization, if $\alpha'$ is replaced by the fine structure constant of the coupling by which
the charged particles reach thermal equilibrium. 
If the charged particles thermalize, the conductivity would be roughly given by~\cite{Baym:1997gq}
\begin{equation}
\sigma_c \sim 10^2 T \sim 10^2 \rho_\rad^{1/4}
\approx5\times 10^{15}\GeV  \left(\frac{\Gamma_\phi}{10^7\GeV}\right)^{1/4}
\left(\frac{H}{10^{11}\GeV}\right)^{1/4},
\quad({\rm thermalized}).
\end{equation}
Hence $\sigma_c$ would be much higher than the Hubble at the decoupling between the inflaton and the gauge field, $H_\dec\sim 10^{11}\GeV,$ and
the generation of the helical magnetic fields would be highly suppressed.
However, this estimation is invalid unless the interaction which brings the decay products into thermal equilibrium is effective, $\Gamma_{\rm int}\gtrsim H$. Thus, eq.~\eqref{case 1} gives the relevant constraint on $\Gamma_\phi$ as
\begin{equation}
\Gamma_\phi \ll 10^{7}\GeV  \left(\frac{\alpha'}{0.01}\right)^{-2}.
\end{equation}
Hence we substitute $\Gamma_\phi=10^6\GeV$ in eq.~\eqref{Beff result}.

\subsection{Case 2 : the inflaton decays into non-charged particles}

Second, we consider the case where the inflaton mainly decays into 
non-charged particles with a constant $\Gamma_\phi$, 
while the non-charged particles can produce
charged particles through an interaction with a cross section, $\sigma_{\rm int} \simeq \tilde{\alpha}^2/p^2$.
In this case, since the energy density of the charged particles is additionally suppressed by factor of $\tilde{\Gamma}_{\rm int}/H\simeq \tilde{\alpha}^2 \Gamma_\phi\Mpl^2/m^3$, the interaction rate between the charged particles and the gauge field is given by
\begin{equation}
\frac{\Gamma_{\rm int}}{H} =\frac{n_{\rm c} \sigma_{\rm int} v}{H}
\simeq \left(\alpha'\tilde{\alpha} \frac{\Gamma_\phi \Mpl^2}{m^3}\right)^2
\approx \left(\frac{\alpha'}{0.01}\right)^2 \left(\frac{\tilde{\alpha}}{0.01}\right)^2
\left(\frac{\Gamma_\phi}{10^7\GeV}\right)^2 \left(\frac{m}{2\times10^{13}\GeV}\right)^{-6}.
\label{case2 result}
\end{equation}
The upper bound on $\Gamma_\phi$ is relaxed,
since its dependence is squared.

Provided that the inflaton $\phi$ is an axion, 
namely as in the natural inflation case~\cite{Freese:1990rb}, it mainly decays into gauge fields and the decay rate is written as~\cite{Pajer:2013fsa}
\begin{equation}
\Gamma_{\phi\to AA} = \frac{\alpha^2m^3}{64\pi f^2} \approx 440\GeV
\left(\frac{\alpha\Mpl}{8f}\right)^2\left(\frac{m}{2\times10^{13}\GeV}\right)^{3},
\end{equation}
where we consider the decay into the $U(1)$ gauge field for simplicity, while the axion may decay into other non-Abelian fields.
Decays into fermions are further suppressed due to the helicity by a factor $m_\psi^2/m^2$,
with a fermion mass $m_\psi<m$~\cite{Pajer:2013fsa}.
Since the $U(1)$ gauge field does not have a self-coupling, 
eq.~\eqref{case2 result} is applicable to this case.
Then, charged particles are unlikely to affect 
the dynamics of the $U(1)$ gauge field calculated in sec.~\ref{Numerical Result}. However, with such a small decay rate,
the inflaton oscillation phase lasts for a long period and
the produced magnetic fields substantially dilute as $B_{\rm phy}(t)\propto a^{-2}$, until the reheating completes and the inverse cascade process becomes effective.

\subsection{Case 3 : the inflaton decays via Yukawa coupling}

Finally let us discuss the case in which the inflaton mainly decays through a Yukawa coupling,
$y\phi\bar{\psi}\psi$, where $y$ is the coupling constant, and $\psi$ is a fermion. Although the perturbative decay rate of this coupling is $\Gamma_y=y^2 m/8\pi$, the actual decay process is non-trivial.
First, the inflaton field value gives a large effective mass to the fermion, $m_\psi=y\phi$, and the fermion is heavier than the inflaton until $\phi$
sufficiently damps. Second, during the $\phi$ oscillation, the fermion
can be produced when the inflaton passes through its origin, $\phi\simeq 0$, by  the so-called {\it fermionic preheating}~\cite{Greene:1998nh, Peloso:2000hy}.
Although the non-perturbative fermion production is studied in the literature, its backreaction to the inflaton is non-trivial in this process.
Therefore it is beyond the scope of this paper to precisely calculate the maximum $\Gamma_{\phi}$
in the case of Yukawa coupling.

However, as a trial, 
let us estimate the decay rate by ignoring the fermion preheating.
Provided the inflaton does not decay into the fermion until its oscillation
amplitude $\phi_\osc$ decreases to $m/2y$ (namely $m_\psi=m/2$), an optimized Yukawa coupling $y$ is obtained as
\begin{equation}
\Gamma_y = H(\phi_\osc=m/2y)
\quad\Longrightarrow\quad
y=\left(\frac{4\pi m}{\sqrt{6}\Mpl}\right)^{1/3}\approx 0.034 \,
\left( \frac{m}{2 \times 10^{13} \GeV} \right)^{1/3} ,
\label{y value}
\end{equation}
where we have used $H\simeq m\phi_\osc/\sqrt{6}\Mpl$. 
Note if $y$ is larger than this value, reheating is delayed because 
the fermion effective mass becomes larger and prevents the decay.
The coupling constant in eq.~\eqref{y value} corresponds to
the decay rate $\Gamma_y \approx 10^9\GeV$.
Therefore it would be interesting to further study the Yukawa coupling case,
since it may realize a high reheating temperature avoiding 
charged particles from suppressing the magnetic field production.

\section{The inflaton perturbation $\delta\phi$}
\label{The inlfaton perturbation}

In sec.~\ref{Numerical Result}, we focus on the case of $\alpha\Mpl/f=8$.
One expects if the coupling constant is larger, stronger magnetic fields are produced. However, at the same time, one should care about the consistency of the assumption, $\delta\phi\ll\phi_0$, which we made in the equation of motion for the gauge field, eq.~\eqref{EoM gauge sim}. If the coupling between the inflaton and the gauge field
is too strong, the perturbation of the inflaton $\delta\phi$ is significantly produced by the gauge field and $\delta\phi$ becomes non-negligible compared to $\phi_0$.
Then $\delta\phi$ potentially alters the production of the gauge field,
while it is difficult to take into account the non-linear coupling
between $\delta\phi$ and $\mcA_\pm$ without performing full lattice simulations.
To find the allowed maximum value of the coupling constant where our assumption
is valid, we evaluate the variance of the inflaton perturbation $\langle\delta\phi^2\rangle$
by ignoring the backreaction from $\delta\phi$ to the gauge field.
That ``maximum" coupling constant leads to the strongest
magnetic field which is obtained under our assumption.

The equation of motion for $\delta\phi(\eta,\bm{x})$ is given by
\begin{equation}
\left\{ \partial_\eta^2 -\partial_i^2 - \frac{a''}{a}
+a^2 \left[ m^2 +2 \frac{m^2\phi_0\dot{\phi_0}}{\Mpl^2 H}
+\left(3+\frac{\dot{H}}{H^2}\right) \frac{\dot{\phi}^2_0}{\Mpl^2}\right]\right\}
\left(a\delta\phi\right)=
a^3 \frac{\alpha}{f} \Big(\bm{E}\cdot\bmB-\langle\bm{E}\cdot\bmB\rangle\Big),
\label{EoM for perturbation}
\end{equation}
where the gravitational coupling to the background inflaton $\phi_0$ is also taken into account.%
\footnote{Since these gravitational coupling terms cause the metric preheating of $\delta\phi$~\cite{Kodama:1996jh, Nambu:1996gf, Finelli:1998bu} and the oscillation amplitude of $\delta\phi$ does not decrease while $\phi_0$
damps as $a^{-3/2}$ after the decoupling. Thus $\delta\phi$ may become non-negligible
eventually and contribute to magnetogenesis if the metric preheating lasts sufficiently long. We will come back to  this issue in a future work.}
The right hand side of eq.~\eqref{EoM for perturbation} represents the contribution of the gauge field through the coupling.
The solution for the Fourier mode of $\delta\phi(\eta,\bm{x})$ coming from the source can be found by using the Green function as
\begin{equation}
a(\eta)\delta\phi(\bm{k},\eta)=
2\int^\eta \dd\tau\, {\rm Im}\big[Q^*(k,\eta)Q(k,\tau) \big]\, J_\em(\bm{k},\tau),
\end{equation}
where $Q(k,\eta)$ is the homogeneous solution of eq.~\eqref{EoM for perturbation},
and $J_\em$ is the Fourier-transformed source term,
\begin{equation}
J_\em (\bm{k},\eta)\equiv a^3(\eta)\frac{\alpha}{f} \int
\frac{\dd^3p}{(2\pi)^3} E_i(\bm{p},\eta)B_i(\bm{k-p},\eta).
\end{equation}
After a few lines of algebra, we obtain the power spectrum of the $\delta\phi$
induced by the gauge field as
\begin{align}
&
\mcP_{\delta\phi}(k,\eta)=
\frac{\alpha^2 k^3}{2\pi^2f^2a^2(\eta)}
\sum_{\lambda,\sigma=\pm}\int \frac{\dd^3 p}{(2\pi)^3} 
\left(1-\lambda\sigma\widehat{\bm{p}}\cdot\widehat{\bm{k-p}}\right)^2
\notag\\
&\times \left[ p^2 \left|\mathcal{I}_{\lambda\sigma}\left(\tau,k;p,|\bm{k-p}|\right) \right|^2 + \lambda\sigma p |\bm{k}-\bm{p}|  \mathcal{I}_{\lambda\sigma}\left(\tau,k;p,|\bm{k-p}|\right)
\mathcal{I}_{\sigma\lambda}^*\left(\tau,k;|\bm{k-p}|,p\right)\right],
\end{align}
with
\begin{equation}
\mathcal{I}_{\lambda\sigma}\left(\eta,k;p,q\right)=
\int^\eta \frac{\dd\tau}{a(\tau)}\, {\rm Im}\big[ Q(k,\eta)Q^*(k,\tau)\big]
\,\mcA_\lambda(p,\tau)\mcA_\sigma'(q,\tau).
\end{equation}
The variance of the inflaton perturbation is given by
$\langle\delta\phi^2(\eta)\rangle=\int\dd k \mcP_{\delta\phi}(k,\eta)/k.$
%
\begin{figure}[tbp]
  \hspace{-2mm}
  \includegraphics[width=75mm]{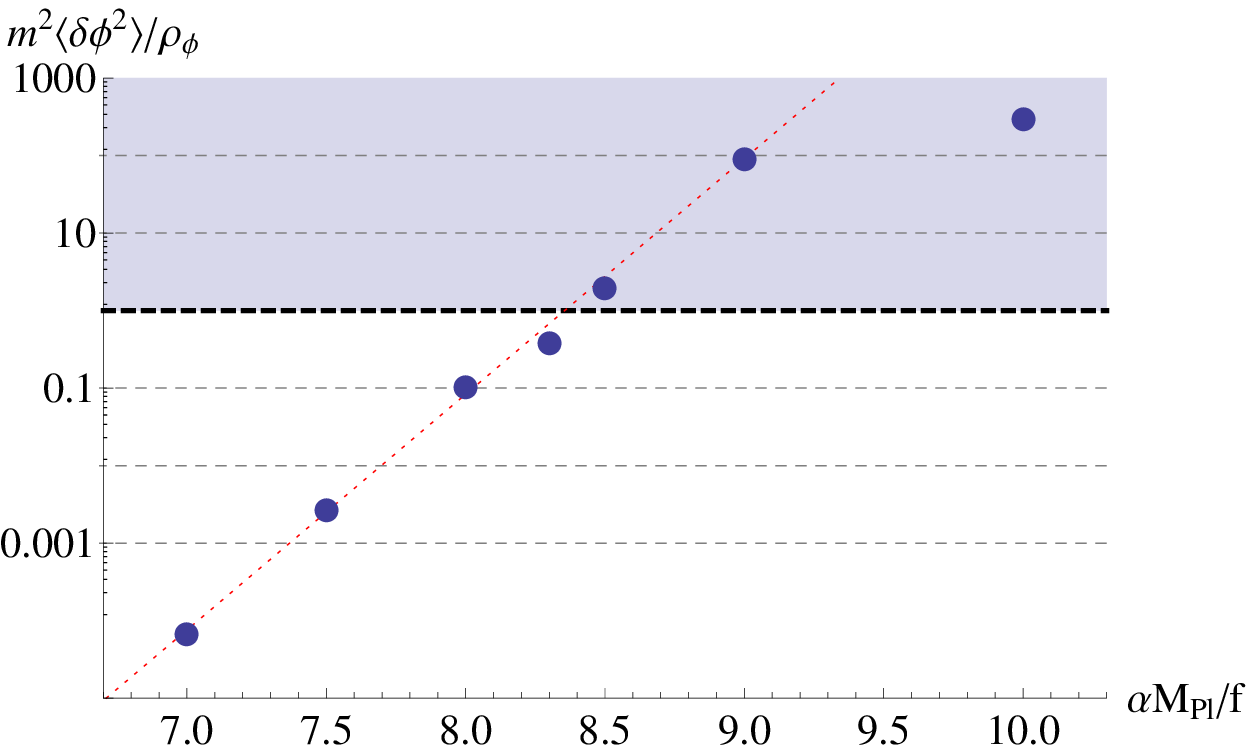}
  \hspace{5mm}
  \includegraphics[width=75mm]{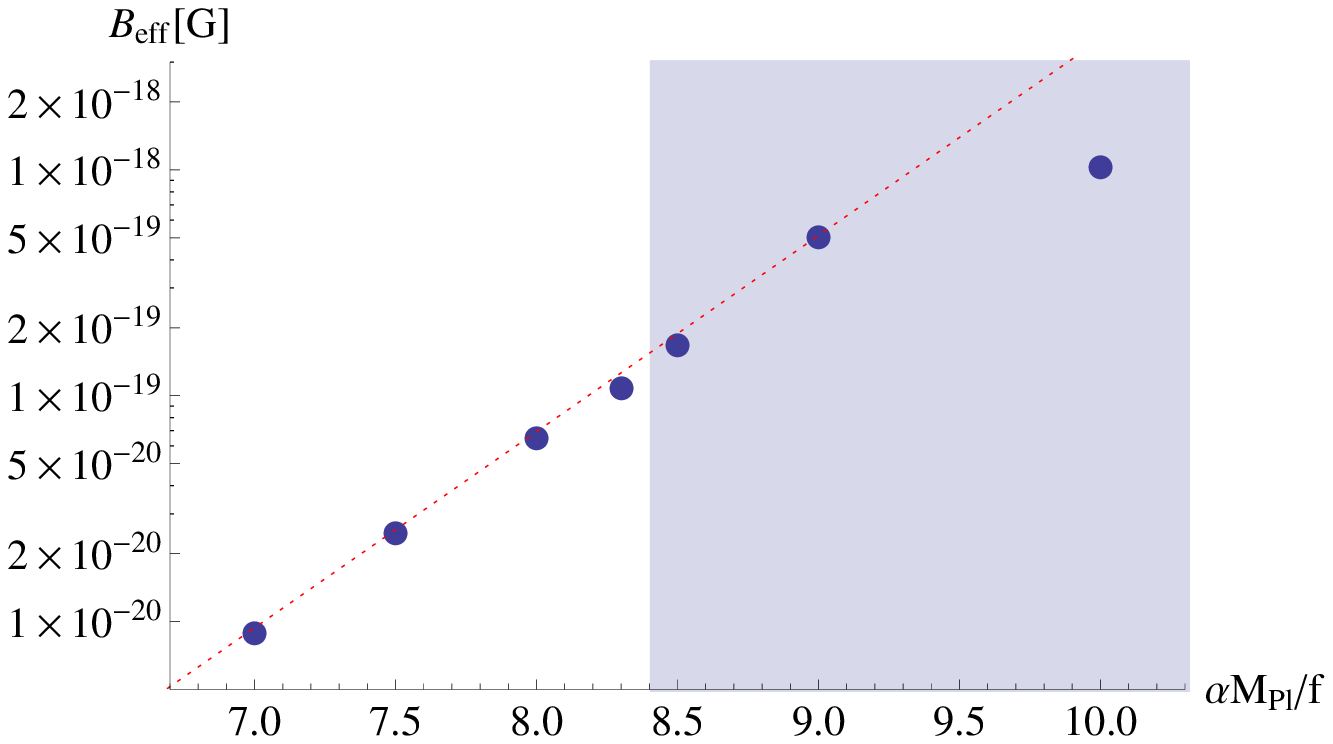}
  \caption
 { The backreaction measure $m^2 \langle \delta\phi^2\rangle/\rho_\phi$
at the time when $\rho_\em/\rho_\phi$ reaches its maximum value (left panel),
and the effective magnetic field strength $B_{\rm eff}$ with $\Gamma_\phi = 10^{6}\GeV$ (right panel) are shown as functions of the coupling constant. Since $\rho_\phi/m^2$ represents the cycle average of $\phi_0^2(t)$, $m^2 \langle \delta\phi^2\rangle/\rho_\phi$ should be smaller than unity to ignore $\delta\phi$ in eq.~\eqref{EoM gauge sim}.
This consistency condition ($\alpha\Mpl/f <8.4$) restricts the reliable prediction for $B_{\rm eff}$ in our framework as $B_{\rm eff}\lesssim 10^{-19}\G\left(\Gamma_\phi/10^6 \GeV\right)^{1/4}$. 
The fitted lines (red dotted) in the left and right panels are given in eqs.~\eqref{fitting line-dphi} and \eqref{fitting line}, respectively.}
 \label{coupling figure}
\end{figure}
%
We numerically evaluate $\langle \delta\phi^2\rangle$ by using 
the solutions of $\phi_0(t)$ and $\mcA_\pm(k,t)$ which are also numerically obtained in sec.~\ref{Numerical Result}.
We compare $\langle \delta\phi^2\rangle$ with $\rho_\phi/m^2$,
which gives the cycle average of 
$\phi_0^2$.

In the left panel of fig.~\ref{coupling figure}, we show $m^2 \langle \delta\phi^2\rangle/\rho_\phi$
at the time when the ratio between the energy density of the gauge field
and that of the background inflaton, namely $\rho_\em/\rho_\phi$, reaches
its maximum value for various values of the coupling constant, $\alpha\Mpl/f =7, 7.5, 8, 8.3, 8.5, 9$ and $10$.
For example, in the case of $\alpha\Mpl/f=8$, $\langle \delta\phi^2\rangle$ plotted in fig.~\ref{coupling figure} is evaluated at $N\approx 2$.
For $\alpha\Mpl/f\ge8.4$, the inflaton perturbation $\delta\phi$ becomes larger than the background value $\phi_0$ before the production of the magnetic fields terminates. Therefore our treatment is justified only for $\alpha\Mpl/f<8.4$. For $\alpha\Mpl/f=8.3$, we find $m^2 \langle \delta\phi^2\rangle/\rho_\phi=0.2$
when $\rho_\em/\rho_\phi$ reaches its maximum and then we obtain 
\begin{equation}
B_{\rm eff}= 1.1\times 10^{-19}\G \left(\frac{\Gamma_\phi}{10^6 \GeV}\right)^{1/4},
\qquad
\left( \frac{\alpha \Mpl}{f}=8.3\right).
\label{Beffmax8.3}
\end{equation}
Therefore this is the strongest magnetic field in our treatment.
In the right panel of fig.~\ref{coupling figure}, $B_{\rm eff}$ at present is shown as a function of the coupling constant, $\alpha\Mpl/f$.
One can clearly see that the resultant  magnetic field becomes stronger as the coupling is larger. 

We find that $m^2 \langle \delta\phi^2\rangle/\rho_\phi$ and $B_{\rm eff}$ are well approximated by the following function (the red dotted lines in  fig.~\ref{coupling figure}):
\begin{align}
&\frac{m^2 \langle \delta\phi^2\rangle}{\rho_\phi}
=4\times 10^{-26}\exp[7\alpha\Mpl/f\big],
&\left( \frac{\alpha \Mpl}{f}\lesssim9\right),
\label{fitting line-dphi}
\\
&B_{\rm eff} = 8\times 10^{-27}\G\, \exp\big[2\alpha\Mpl/f\big]\left(\frac{\Gamma_\phi}{10^6 \GeV}\right)^{1/4},
& \left( \frac{\alpha \Mpl}{f}\lesssim9\right).
\label{fitting line}
\end{align}
For $\alpha\Mpl/f\gtrsim9$, where our computation is invalid as $\delta\phi > \phi_0$, $\rho_\em$ becomes much larger than $\rho_\phi$ in our numerical calculation,
not only because the amplification from the tachyonic instability gets stronger but also because the backreaction from the gauge field suppresses the amplitude of the inflaton. Since $\rho_\em \gg \rho_\phi$, the universe is radiation dominated for a while after the decoupling, and the dilution of the produced magnetic fields is less significant. Accordingly, $B_{\rm eff}$ becomes stronger as the coupling is larger, although the almost all energy of the inflaton
is transferred into the gauge field for $\alpha\Mpl/f\gtrsim9$. However, such a large coupling constant invalidates the assumptions in our calculation, and the result is likely to be modified, at least quantitatively, if $\delta\phi$ is properly taken into account.
\footnote{
In the recent paper~\cite{Adshead:2015pva}, the coupled dynamics of the inflaton and gauge field is studied by lattice simulations for larger coupling cases. The simulations with  analytically approximated initial conditions of inflaton and gauge field at the end of inflation show that the helical asymmetry of the gauge field would be erased 
by the re-scattering process for $\alpha/f\gtrsim9M_p^{-1}$. 
However, the homogenous mode of the inflaton and the configuration of the gauge field deviates from the analytical form for the larger axial coupling as mentioned in~\cite{Adshead:2015pva},
which has the possibility to change the eventual helical asymmetry.
Thus, in order to evaluate the abundance of the magnetic field for larger coupling at least around $\alpha/f\simeq 9M_p^{-1}$, one needs to execute the non-linear simulations such as lattice simulations to take into account the re-scattering process with appropriate initial conditions.
}

The result, eq.~\eqref{Beffmax8.3}, for $\alpha \Mpl/f=8.3$ and $\Gamma_\phi =10^6\GeV$ corresponds to the magnetic fields with the strength $B_{\rm phy}(t_{\rm now}) \sim 5\times10^{-16}\G$ and the correlation length $\lambda_{\rm phy}(t_{\rm now})\sim 0.05$pc.
Thus, in spite of the significant growth due to the inverse cascade,
the correlation length of the magnetic field is still very small.
It is because we have considered the large-field model of inflation 
in which the energy scale around the end of inflation is very high 
and the magnetic fields are generated on a very small scale.
Note that the mass scale $m=1.9\times 10^{13}\GeV$ corresponds
to the length scale $m^{-1}=3\times10^{-52}\Mpc$.

In order to produce magnetic fields with a larger correlation length,
one can consider a lower-energy inflation model or a spectator field
with the axial coupling which begins to oscillate at a lower-energy scale after inflation.
Although such a model may be less simple and contain more parameters than
our setup in this paper, it is expected that one can calculate resultant magnetic fields in a similar manner. We intend to investigate this possibility in the future work.

A spectator field with a generalized axial coupling in a low energy inflation has been considered in refs.~\cite{Durrer:2010mq, Caprini:2014mja}.
In these works, the authors assumed that $\xi$ is constant until instantaneous reheating without solving the dynamics of the spectator field. 
In this paper, however, we have shown that $\xi$ can vary in a non-trivial way and it leads to an efficient amplification
of magnetic fields. We have also discussed the  effects of charged particles and the perturbation of the axial coupled scalar field.
It is also worth reconsidering these models taking them into account.

\section{Conclusion}
\label{Conclusion}

In this paper, we investigate the generation of magnetic fields in the axial coupling model where the inflaton and the gauge field are coupled through the term, ${\cal L}_{\rm int} = - \frac{\alpha}{4f}\phi F\tilde{F}$.
Although only the slow-roll regime of the inflaton is considered in the previous works, we point out, for the first time in the context of magnetogenesis, that the most efficient production of the electromagnetic fields takes place after the end of inflation.
This property of the model is quite interesting because it is known
that magnetogenesis only during inflation suffers from several problems
and thus the production of magnetic field with the sufficient strength to
explain the blazar observations is difficult. 

We have numerically solved the coupled equations of the background inflaton and the gauge field by taking into account the backreaction from the produced gauge field to the inflaton and the background dynamics.
Since the axial coupling spontaneously breaks the parity symmetry, 
one of the two polarizations of the gauge field is mainly produced.
In our case where $\dot\phi$ is initially negative, only the $(-)$ mode, $\mcA_-(k,t)$, acquires the double amplification
from the tachyonic instability and the parametric resonance.
For the opposite signature of $\dot\phi$, only the $(+)$ mode enhances.
As a result, the almost completely helical magnetic field is produced.  
After turbulence of the magnetized plasma develops, the helical magnetic fields undergo the inverse cascade process, and their comoving amplitude
grows. In virtue of the helicity conservation, we can easily calculate 
the effective magnetic strength $B_{\rm eff}$ at present from the physical
intensity and correlation length of the magnetic fields at reheating.

In this paper, we make two assumptions to simplify the calculation.
We assume that electrically charged particles produced by the inflaton decay
do not significantly interact with the gauge field until the production of the magnetic fields  terminates, and that the coupling constant of the axial coupling is small enough to ignore the effect of the inflaton perturbation $\delta\phi$ on the gauge field dynamics.  
Consequently, we find that the maximum value of $B_{\rm eff}$ in our framework is $10^{-19}\G$ and it is not sufficient to explain the Blazar observations.
However, the above two assumptions can be relaxed,
for example, by phenomenologically taking into account the effect of charged particles~\cite{Bassett:2000aw}
and performing lattice simulations, respectively.

We have considered the inflaton with the quadratic mass potential and 
the axial coupling in this paper, because it is one of the simplest setups
to generate magnetic fields in the primordial universe.
Unfortunately, however, we have found that the produced magnetic fields have
a too small correlation length, $\lambda_{\rm phy}(t_{\rm now}) \lesssim 0.1 $pc,
and hence its effective strength is too weak to explain the blazar observations.
Therefore, as a natural extension of this work,
it would be interesting to consider a scenario in which
magnetogenesis due to the axial coupling takes place at a lower energy scale.
For example, in a small field inflation model,
inflation ends at a lower energy and hence magnetic fields are produced
on a larger scale. 
Alternatively, one can also consider that
a spectator (pseudo-)scalar field has an axial coupling to the gauge field.
When the spectator field begins to oscillate during the oscillation phase of the inflaton or (dark) radiation dominated era after inflation, electromagnetic fields can be generated in a similar manner to that found in this paper, if the inflaton decays into a dark sector without electrically charged particles.
Although models of a low energy inflation and/or 
a spectator field
contain more parameters than the simple case explored in this paper, one can compute the resultant strength of the magnetic fields in the same way, once a concrete model is specified.


\acknowledgments

We would like to thank Chiara Caprini, Masahiro Kawasaki, Takeo Moroi, Misao Sasaki and Yuki Watanabe for useful comments.
This work was supported by the World Premier International
Research Center Initiative (WPI Initiative), MEXT, Japan. 
We acknowledge the supports by Grant-in-Aid for JSPS Fellows
No.248160 [T.F.],  the Advanced Leading Graduate Course for
Photon Science grant [Y.T.], MEXT's Program for Leading Graduate Schools
PhD professional,  ``Gateway to Success in Frontier Asia,'' and JSPS Grant-in-Aid
for Scientific Research No. 25287057 and 26887020 [H.T].
This work has been partially done within the Labex ILP (reference ANR-10-LABX-63) part of the Idex SUPER, and received financial state aid managed by the Agence Nationale de la Recherche, as part of the programme Investissements d'avenir under the reference ANR-11-IDEX-0004-02.


\end{document}